# Calculation of Integrals in MathPartner

**Gennadi I. Malaschonok[1], Alexandr V. Seliverstov[2]**

[1] *National University of Kyiv-Mohyla Academy*
*2, Skovorody St., Kyiv, 04070, Ukraine*
[2] *Institute for Information Transmission Problems*
*of the Russian Academy of Sciences (Kharkevich Institute)*
*19-1, Bolshoy Karetny per., Moscow, 127051, Russian Federation*

We present the possibilities provided by the MathPartner service of calculating definite and indefinite integrals. MathPartner contains software implementation of the Risch algorithm and provides users with the ability to compute antiderivatives for elementary functions. Certain integrals, including improper integrals, can be calculated using numerical algorithms. In this case, every user has the ability to indicate the required accuracy with which he needs to know the numerical value of the integral. We highlight special functions allowing us to calculate complete elliptic integrals. These include functions for calculating the arithmetic-geometric mean and the geometric-harmonic mean, which allow us to calculate the complete elliptic integrals of the first kind. The set also includes the modified arithmetic-geometric mean, proposed by Semjon Adlaj, which allows us to calculate the complete elliptic integrals of the second kind as well as the circumference of an ellipse. The Lagutinski algorithm is of particular interest. For given differentiation in the field of bivariate rational functions, one can decide whether there exists a rational integral. The algorithm is based on calculating the Lagutinski determinant. Mikhail Lagutinski (1871–1915) had worked at Kharkiv. This year we are celebrating his 150th anniversary.

## 1. Introduction

The development of computer algebra systems and cloud computing makes it possible to solve many computational problems. Vladimir Petrovich Gerdt was at the forefront of the development of computer algebra. As a professional physicist, he developed new algorithms for solving problems in mathematical physics and implemented them in many well-known systems of computer algebra. He has worked on systems such as REDUCE, Mathematica, Maple, and Singular.

Today, many useful programs and cloud services are available. A new generation of computer algebra systems is actively developing. They are cloud-based systems freely available on the Internet. The MathPartner service is a nice example of this [1]–[4]. Free access to the MathPartner service is possible at `http://mathpar.ukma.edu.ua/` as well as `http://mathpar.com/`.

In this review, we consider only a small area of MathPartner application, namely the calculation of definite and indefinite integrals. Symbolic computations and estimates of the computational complexity are of the greatest interest [5]–[9]. However, in some cases, symbolic computations need to be supplemented with numerical methods. In particular, this is true when calculating special functions [10], [11]. For example, elliptic integrals are used to calculate the period of the simple pendulum [12] as well as some properties of porous materials [13], [14].

Robert Henry Risch proposed a method to integrate elementary functions [15], [16]. The method was later improved by Manuel Bronstein [17]. In 2010–2019, an algorithm based on the Liouville–Risch–Davenport–Trager–Bronstein theory was developed at the Laboratory of Algebraic Computations of Derzhavin Tambov State University. A series of papers on symbolic integration algorithms was published by Svetlana Mikhailovna Tararova [18] and Vyacheslav Alekseevich Korabelnikov [19], [20]. The procedures were developed using object-oriented programming in Java. Their description is given in cited publications. Since the symbolic integration theory has not yet been completed, this algorithm can be considered as a good basis for further theoretical and practical development in this important area.

Historically, the first major symbolic integration project was the IBM Scratchpad project led by Richard Dimick Jenks. The development of this project as a commercial one was later stopped by the company. However, he played an important role in the development of the theory of symbolic integration and attracting interest in it.

Many general computer algebra systems today support symbolic integration of elementary functions. However, they all have a common drawback that is the incompleteness of solving the problem of symbolic integration. Another drawback is the lack of a detailed description of the procedural implementation and the technical possibility of further development of the package of procedures. The most famous example is the cloud-based SAGE system, which provides access to old open source packages that have long been discontinued. On the other hand, commercial systems do not give users access to their packages of procedures, and they do not have specialists who can complete the theory of calculating the antiderivative for the composition of simple elementary functions.

Experiments with integration problems from mathematical analysis textbooks show that many problems can be solved using any of the systems such as Mathematica, Maple, and MathPartner. Nevertheless, for each of them, one can find functions that have an antiderivative, but it is not calculated by this system. The MathPartner symbolic integration package is one of the newest packages in this area. It is developed in Java and is the most promising for further development.

In a series of important works, Mikhail Nikolaevich Lagutinski (1871–1915) developed a method for determining integrals of polynomial ordinary differen-tial equations in finite terms. He also developed the theory of integrability in finite terms of such systems of equations [21]–[23]. Lagutinski was an outstanding mathematician. He had worked at Kharkiv and died during the First World War. In this article, we also consider the Lagutinski method.

Note that he published his papers as Lagoutinsky using the French spelling [24], [25]. The authors are grateful to Mikhail Malykh for comments and historical notes about M. N. Lagutinski.

## 2. Integrals of some functions

### 2.1. Indefinite integrals

To calculate the indefinite integral of an elementary function $f(x)$ one can run the command $\mathbf{int}(f)dx$, where $x$ is declared in the environment SPACE. Five number sets $\mathbb{Q}$, $\mathbb{R}$, $\mathbb{R}64$, $\mathbb{C}$, and $\mathbb{C}64$ can be used. Over the field $\mathbb{Q}$, pure symbolic computations are done. For example, let us calculate $\int 2x\sin(x^2)\,dx$:

```
SPACE = Q[x];
f = 2*x*\sin(x^2);
\int(f) d x;
```

The output is equal to $(-1)\cos(x^2)$.

Next, let us calculate $\int (3x^2+2)^2\,dx$:

```
SPACE = Q[x];
\int((3x^2+2)^2) d x;
```

The output is equal to $(9/5)x^5 + 4x^3 + 4x$.

### 2.2. Definite integrals (the numerical algorithm)

A definite integral $\int_a^b f(x)\,dx$ can be calculated by means of the command $\mathbf{Nint}(f, a, b, \varepsilon, N)$, where $\varepsilon$ means the approximation to $\varepsilon$ decimals and $N$ denotes the number of points in the Gaussian formula (optional). These parameters can be omitted.

For example, let us calculate 42 decimal places of the integral $\int_0^\pi \sin x\,dx$:

```
SPACE = R[x]; MachineEpsilonR = 42; FLOATPOS = 42;
\Nint(\sin(x), 0, \pi);
```

The output is equal to 2.000000000000000000000000000000000000000000.

Next, let us approximate $\pi$:

```
SPACE = R[x]; MachineEpsilonR = 43; FLOATPOS = 42;
2*\Nint(\sqrt(1-x^2), -1,1);
```

The output is equal to 3.141592653589793238462643383279502884197169. All 42 decimal places are accurate.

Let us consider improper integrals of the first kind $\int_a^b f(t)\,dt$, where $a$ or $b$ can be equal to $\pm\infty$. The integral can be calculated by means of the command $\mathbf{Nint}(f, a, b, [q_1, \cdots, q_m], \varepsilon, N)$, where $[q_1, \cdots, q_m]$ denotes the set of extreme points of the function $f$ inside the interval of integration $(a, b)$, the answer is the approximation to $\varepsilon$ decimals, and $N$ denotes the number of points in the Gaussian formula (optional). In fact, three parameters $[q_1, \cdots, q_m]$, $\varepsilon$, and $N$ can be omitted. If the extreme points are not indicated, then the correctness of the output is ensured when the integrand is monotonic on the interval of integration.

For example, let us calculate $\int_{-\infty}^{+\infty} \exp\left\{-(x-5)^2\right\}\,dx$:

```
SPACE = R64[x];
f = \exp\{-(x-5)^2\};
\Nint(f, -\infty, \infty);
```

The output is equal to 1.77.

Next, let us calculate $\int_0^{\infty} \exp\{-x\}\,dx$:

```
SPACE = R[x];
MachineEpsilonR = 45; FLOATPOS = 45;
\Nint(\exp(-x), 0, \infty);
```

The output is equal to 1.000000000000000000000000000000000000000000000, where all 45 decimal places are accurate.

### 2.3. The complete elliptic integrals

For some improper integrals, more efficient calculation methods are known. Let us consider complete elliptic integrals [10], [12]. For positive numbers $a > 0$ and $b > 0$, the complete elliptic integral of the first kind can be calculated by means of the arithmetic-geometric mean

$$I(a,b) = \int_0^{\pi/2} \frac{d\varphi}{\sqrt{a^2\cos^2\varphi + b^2\sin^2\varphi}} = \frac{\pi}{2\mathbf{AGM}(a,b)},$$

where $\mathbf{AGM}(a,b)$ denotes the arithmetic-geometric mean of $a$ and $b$. It is equal to the limit of both sequences $a_n$ and $b_n$, where $a_0 = a$, $b_0 = b$, $a_{n+1} = \frac{1}{2}(a_n + b_n)$, and $b_{n+1} = \sqrt{a_n b_n}$. The proof is based on the equality $I(a,b) = I\left((a+b)/2, \sqrt{ab}\right)$. Of course, if $a = b$, then $I(a,a) = \pi/2a$.

On the other hand, the geometric-harmonic mean $\mathbf{GHM}(a,b)$ is equal to the limit of both sequences $a_n$ and $b_n$, where $a_0 = a$, $b_0 = b$, $a_{n+1} = \sqrt{a_n b_n}$, and $b_{n+1} = \frac{2a_n b_n}{a_n + b_n}$. Note that $\mathbf{AGM}(a,b)\mathbf{GHM}(a,b) = ab$.

Both $\mathbf{AGM}(a,b)$ and $\mathbf{GHM}(a,b)$ can be calculated in the MathPartner service. For example, let us run the commands, where FLOATPOS denotes the number of decimal places:

```
SPACE = R64[];
FLOATPOS = 3;
a= \AGM(1, 5);
g= \GHM(1, 5);
\print(a, g);
```

The output is $a = 2.604$ and $g = 1.920$.

The complete elliptic integral of the second kind can be calculated by means of **MAGM**(). It is also implemented in the MathPartner service. The modified arithmetic-geometric mean $\mathbf{MAGM}(a, b)$ is equal to the limit of the sequence $a_n$, where $a_0 = a$, $b_0 = b$, $c_0 = 0$, $a_{n+1} = a_n + b_n/2$, $b_{n+1} = c_n + \sqrt{(a_n - c_n)(b_n - c_n)}$, and $c_{n+1} = c_n - \sqrt{(a_n - c_n)(b_n - c_n)}$,

$$\int_0^{\pi/2} \sqrt{a^2 \cos^2 \varphi + b^2 \sin^2 \varphi}\, d\varphi = \frac{\pi}{2} \cdot \frac{\mathrm{MAGM}(a^2, b^2)}{\mathrm{AGM}(a, b)}.$$

So, the circumference of an ellipse is equal to $2\pi \dfrac{\mathbf{MAGM}(a^2, b^2)}{\mathbf{AGM}(a, b)}$, where $a$ and $b$ denote the semi-major and semi-minor axes.

## 2.4. Other special functions

The gamma function is defined via a convergent improper integral

$$\Gamma(z) = \int_0^{\infty} x^{z-1} \exp\{-x\}\, dx,$$

where $\mathrm{Re}\, z > 0$. For any positive integer $n$, $\Gamma(n) = (n-1)!$. To calculate its value one can run the command $\mathbf{Gamma}(z)$.

The beta function, also called the Euler integral of the first kind, is also defined via another integral

$$\mathrm{B}(x, y) = \int_0^1 t^{x-1}(1-t)^{y-1}\, dt,$$

where both inequalities hold $\mathrm{Re}\, x > 0$ and $\mathrm{Re}\, y > 0$. It is closely related to the gamma function because $\mathrm{B}(x, y) = \Gamma(x)\Gamma(y)/\Gamma(x+y)$.

To calculate its value one can run the command $\mathbf{Beta}(x, y)$. For example, let us verify the equality $B(2, 3) = 1/12$:

```
SPACE = R64[];
FLOATPOS = 4;
\Beta(2, 3);
```

The output is equal to 0.0833.

The binomial coefficients are $\mathbf{binom}(n, k) = \begin{pmatrix} n \\ k \end{pmatrix} = \dfrac{n!}{k!(n-k)!}$. For suitable function $f$, the Laplace transform is the integral $\displaystyle\int_0^{\infty} f(t) e^{-st}\, dt$.

To calculate the Laplace transform one can run **laplaceTransform**(). The inverse Laplace transform can be calculated by **inverseLaplaceTransform**(). Let us consider an example:

```
SPACE = R64[t];
L = \laplaceTransform(\exp(3t));
```

The output is $L = \dfrac{1.0}{t - 3.0}$.

Next, let us calculate the inverse transform:

```
SPACE = R64[t];
F = \inverseLaplaceTransform(1/(t - 3));
```

The output is $F = e^{3t}$.

## 3. The Lagutinski determinant

The Lagutinski method allows us to search for rational integrals of a given differential ring [21], [24], [25]. Therefore, it can be used to integrate ordinary differential equations in symbolic form [7], [23].

Let us consider the differential ring $\mathbb{Q}[x, y]$ of bivariate polynomials over the field $\mathbb{Q}$, where the differentiation is given by $D = p(x, y)\dfrac{\partial}{\partial x} + q(x, y)\dfrac{\partial}{\partial y}$.

Let us consider an infinite matrix whose entries are monomials. The first row consists of all bivariate monomials with graduated lexicographical ordering $m_1$, $m_2$, .... The second row consists of the first derivatives $Dm_1$, $Dm_2$, .... The third row consists of the second derivatives $D^2 m_1$, $D^2 m_2$, ..., and so on. In particular, both monomials $m_2$ and $m_3$ are linear. For $N = \dfrac{1}{2}(d+1)(d+2)$, the monomial $m_N$ is the last monomial of degree $d$.

The Lagutinski determinant of order $n$ with respect to the differentiation $D$ is a leading principal minor of order $n$ in this matrix. Of course, the first order Lagutinski determinant is equal to 1. To calculate the Lagutinski determinant of order $n$ with respect to the differentiation $D$ one can run the command $\det \mathbf{L}(n, [p, x, q, y])$.

The significance of this determinant is explained by the following result that was previously obtained by Lagutinski [24], [25], but presented here in a modern formulation, cf. [7], [23]. A non-constant rational function $f \in \mathbb{Q}(x, y)$ is called an integral when $Df$ vanishes identically.

**Theorem 1 (Lagutinski).** *Given a differentiation $D$ and a positive integer $d > 0$. The Lagutinski determinant of order $N = \dfrac{1}{2}(d + 1)(d + 2)$ vanishes if and only if there exists a rational integral whose numerator and denominator are of degree at most $d$.*

For example, if the differentiation is given by $D = \dfrac{\partial}{\partial x} + \dfrac{\partial}{\partial y}$, then $D(x - y) = 0$. In accordance with Theorem 1, the third order Lagutinski determinant vanishes:

```
SPACE = Q[x, y];
p = 1; q = 1;
\detL(3, [p, x, q, y]);
```

The output is equal to zero.

Next, if the differentiation is given by $D = x\frac{\partial}{\partial x} - y\frac{\partial}{\partial y}$, then $D(xy) = 0$. In accordance with Theorem 1, the sixth order Lagutinski determinant vanishes. Moreover, the fifth order Lagutinski determinant vanishes too because the fifth monomial coincides with the integral:

```
SPACE = Q[x, y];
p = x; q = -y;
\detL(5, [p, x, q, y]);
```

The output is equal to zero. But the third order Lagutinski determinant is equal to $-2xy$. Thus, there does not exist any integral whose numerator and denominator are linear. Contrariwise, if the differentiation is given by $D = x\frac{\partial}{\partial x} + y\frac{\partial}{\partial y}$, then $D\left(\frac{x-y}{x+y}\right) = 0$. In accordance with Theorem 1, the third order Lagutinski determinant vanishes:

```
SPACE = Q[x, y];
p = x; q = y;
\detL(3, [p, x, q, y]);
```

The output is equal to zero.

## 4. Conclusion

The MathPartner service has become better and allows us to solve new problems in geometry and physics. MathPartner supports both symbolic and numerical integration of elementary functions. Moreover, some special functions can be calculated using fast algorithms.

**Information about the authors**:
**Malaschonok, Gennadi I.** — Doctor of Physical and Mathematical Sciences, Professor, Department of Informatics, National University of Kyiv-Mohyla Academy (NaUKMA) (e-mail: malaschonok@gmail.com, phone: +380444256064, ORCID: https://orcid.org/0000-0002-9698-6374, ResearcherID: F-8856-2015, Scopus Author ID: 14054474400)

**Seliverstov, Alexandr V.** — Candidate of Physical and Mathematical Sciences, Leading researcher, Institute for Information Transmission Problems of the Russian Academy of Sciences (e-mail: slvstv@iitp.ru, phone: +74956943338, ORCID: https://orcid.org/0000-0003-4746-6396, ResearcherID: W-1003-2018, Scopus Author ID: 10439983500)

# Вычисление интегралов в MathPartner

**Г. И. Малашонок[1], А. В. Селиверстов[2]**

[1] *ННациональный университет «Киево-Могилянская академия» ул. Григория Сковороды, д. 2, Киев, 04655, Украина*
[2] *Институт проблем передачи информации им. А. А. Харкевича РАН Большой Каретный пер., д. 19-1, Москваб 127051, Россия*

В статье рассмотрены возможности сервиса MathPartner по вычислению определённых и неопределённых интегралов. MathPartner содержит программную реализацию алгоритма Риша и предоставляет пользователям возможность вычислять первообразные для элементарных функций. Некоторые интегралы, в том числе несобственные, можно вычислить с помощью численных алгоритмов. В этом случае каждый пользователь может указать необходимую точность, с которой ему необходимо знать числовое значение интеграла. Отметим специальные функции, которые позволяют вычислять полные эллиптические интегралы. К ним относятся функции для вычисления арифметико-геометрического среднего и геометрическо-гармонического среднего, которые позволяют вычислять полные эллиптические интегралы первого рода. Набор также включает модифицированное арифметико-геометрическое среднее, которое предложил Семён Адлай, что позволяет вычислять полные эллиптические интегралы второго рода и длину (периметр) эллипса. Особый интерес представляет алгоритм Лагутинского. Для данного дифференцирования в поле рациональных функций от двух переменных можно решить, существует ли рациональный интеграл. Алгоритм основан на вычислении определителя Лагутинского. Михаил Лагутинский (1871–1915) работал в Харькове. В этом году мы отмечаем 150-летие со дня его рождения.